\documentclass[british,english,english,twocolumn,showpacs]{revtex4}
\usepackage{float}
\usepackage{graphicx}
\usepackage{amssymb}
\usepackage{dcolumn}
\usepackage{babel}
\makeatother
\usepackage{color}

\begin{document}

\title{Polyhedral colloidal `rocks': low-dimensional networks}

\author{Rebecca Rice$^1$, Roland Roth$^2$ and C. Patrick Royall$^1$}
\affiliation{$^1$School of Chemistry, University of Bristol, Bristol, BS8 1TS, UK}
\affiliation{$^2$Institut f\"ur Theoretische Physik, Universit\"at Erlangen-N\"urnberg, Staudtstr. 7, 91058 Erlangen, Germany.}

\date{\today}

\begin{abstract}
We introduce a model system of anisotropic colloidal `rocks'. Due to their
shape, the bonding introduced via non-absorbing polymers is profoundly
different from spherical particles: bonds between rocks are rigid against
rotation, leading to strong frustration. We develop a geometric model which
captures the essence of the rocks. Experiments and simulations show that the
colloid geometry leads to structures of low fractal dimension. This is in
stark contrast to gels of spheres, whose rigidity results from locally dense
regions. At high density  the rocks form a quasi one-component glass.
\end{abstract}


\maketitle

Dispersions of mesoscopic colloidal particles are important for several reasons. 
They model atomic and molecular systems, yet single particle level imaging reveals local phenomena, such as
pinpointing mechanisms of dynamic arrest \cite{weeks2001,royall2008g},  
which are inaccessible in conventional systems. Furthermore, colloidal and 
nanoparticle systems are important materials. Colloidal gels stabilize a range 
of industrial and consumer products from pesticides to cosmetics. Underlying 
both is the tuneability of interactions between
colloids which are theoretically well understood and hence enable 
design of self-assembled structures at small lengthscales. Recently, 
anisotropic interactions using a number of sticky patches per particle 
have been introduced, opening new routes of self-assembly. Reducing the 
number of patches per particle leads to networks of low fractal dimension
 \cite{bianchi2006}.

\begin{figure*}
\includegraphics[width=40mm]{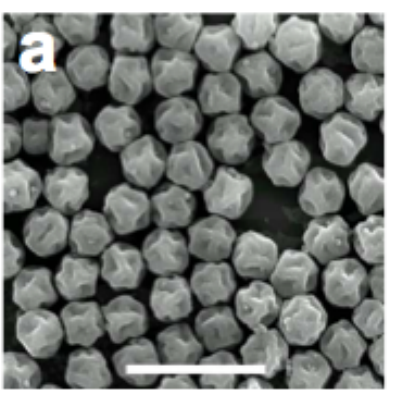}
\includegraphics[width=30mm]{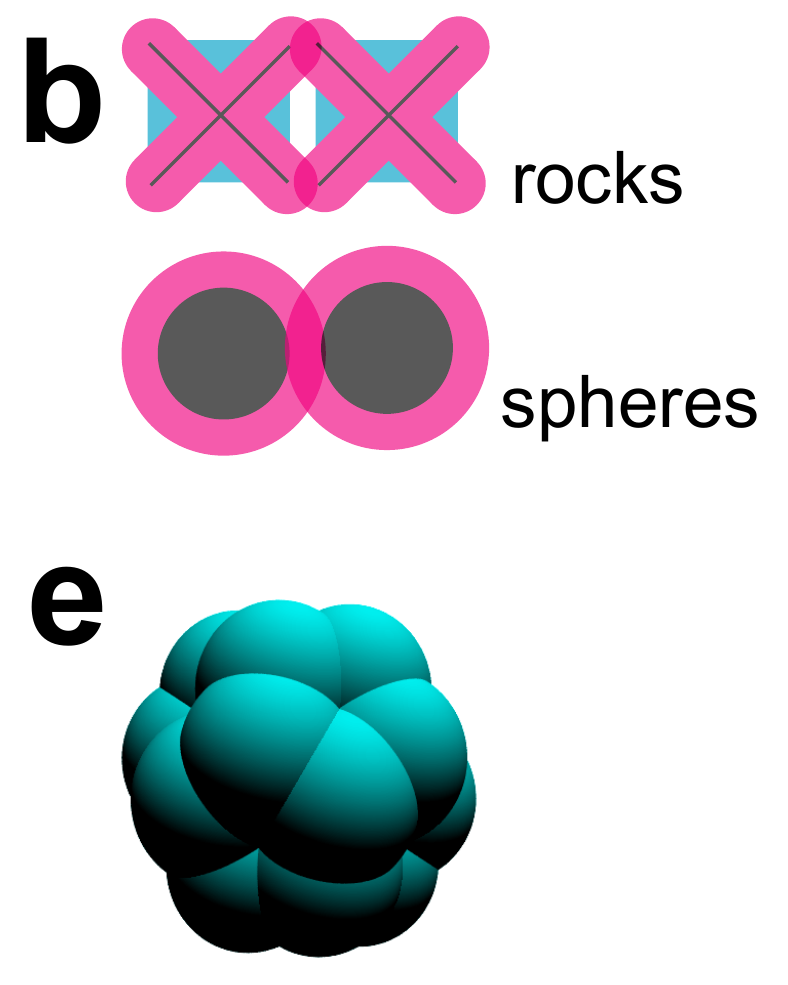}
\includegraphics[width=80mm]{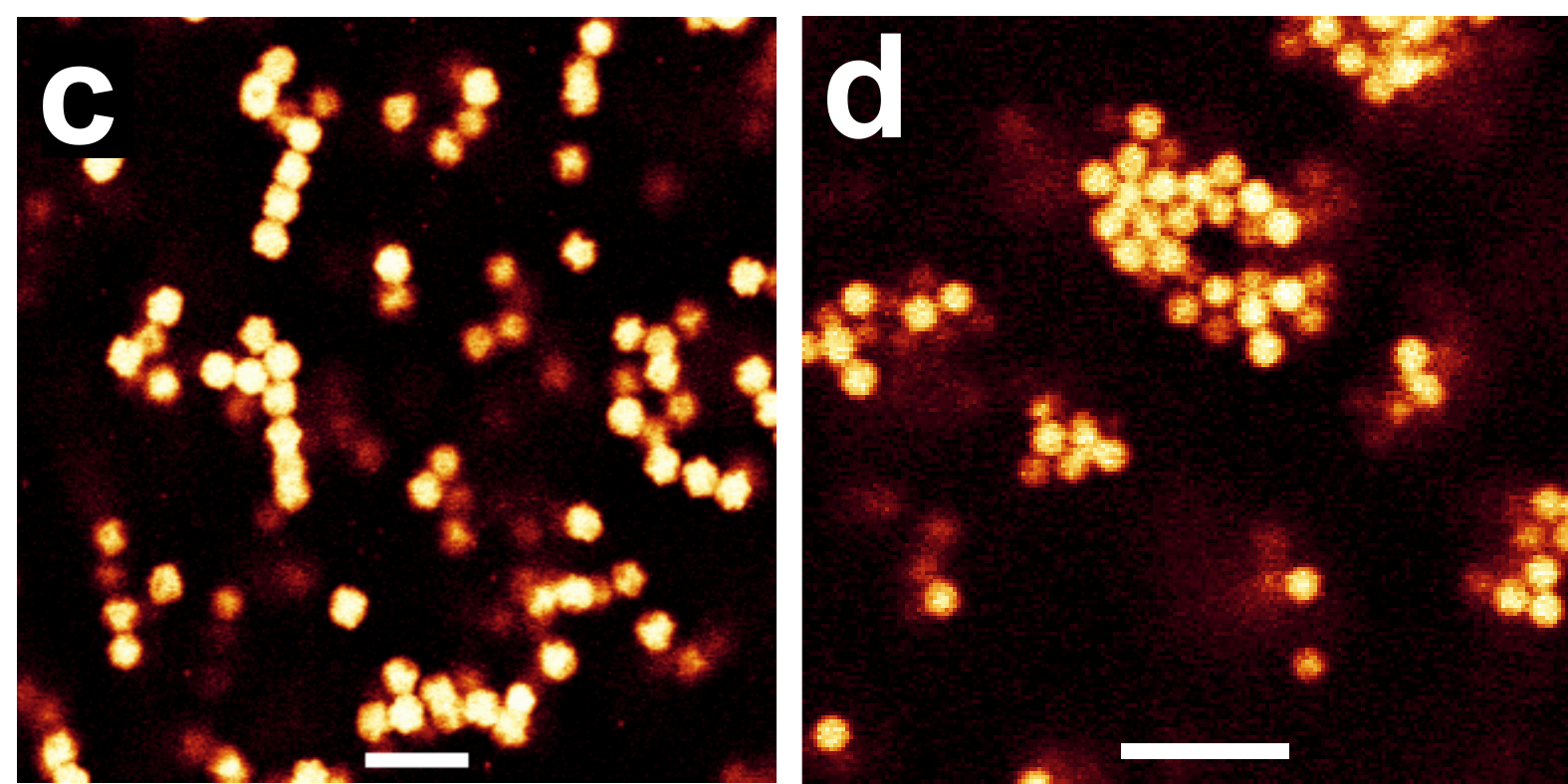}
\caption{(color online) (a) SEM image of colloidal rocks. (b) Schematic of the mechanism
of rigid bonding induced by multiple overlap zones in the rocks contrasted with spheres which are free to rotate. Regions from which the polymer coil center of mass is excluded are shown in pink. (c) Confocal image of a gel of rocks $c_{p}/c_{p}^{gel}=1.13$.
(d) Confocal image of a gel of spheres $c_{p}/c_{p}^{gel}=1.44$. $c_{p}$
is polymer concentration and $c_{p}^{gel}$ is at the onset of gelation, see Fig. \ref{figPerky}. (e) Dodecahedral assembly of 20 spheres used in simulations.
 Scale bars in a,c,d are 10 $\mu$m. 
\label{figPix}}
\end{figure*}

Here we consider depletion induced gels of anisotropic particles. The addition of 
polymer mediates an effective attraction between the
colloids due to excluded volume effects -- see Fig \ref{figPix}(b). The
depth of this attraction is proportional to the polymer concentration $c_p$
and the range is set by the polymer size. At sufficient strengths of
the depletion attraction, spinodal phase separation leads to
colloid-rich (polymer-poor) `colloidal liquid' and colloid-poor
(polymer-rich) `colloidal gas' phases. Decreasing the range of the 
depletion interaction by reducing the size of the polymer relative to the colloid leads
to a higher density colloidal liquid.
If the packing fraction of the colloidal liquid
is high enough ($\phi\approx0.58$) phase 
separation is arrested. The resulting network of voids combined with `arms' of high local
 colloid density, is termed a gel \cite{lu2008}.

A range of anisotropic particles has been synthesized
\cite{glotzer2007,kraft2009}, and systems with hard interactions have recieved
considerable attention very recently: rods form the expected liquid crystalline phases
 \cite{kuijk2011}, anisotropy suppresses long-ranged order close to a wall \cite{dullens2006pot} and at high density, colloidal dumbells show multiple relaxation pathways 
\cite{kramb2010}. However, although networks of anisotropic particles are important, 
e.g. in shiny paper coatings, systematic experimental studies of the effect of anisotropy in
systems with attractions
are limited: attractions between colloidal platelets can lead to stacking \cite{zhao2007,zhang2011}, 
gels of colloidal rods form bundles \cite{wilkins2009} and due to their patchy
interactions, clay platelets form low 
density colloidal `liquids' \cite{ruzicka2010}.  

Here we introduce an
 anisotropic colloidal system of faceted polyhedra, or `rocks', with tuneable 
interactions, amenable to 3D single-particle level analysis with confocal microscopy. 
Unlike gels of spherical particles where phase separation is suppressed by
slow dynamics due to the high local colloid density of the `arms',
we show that the polyhedral
nature of the rocks leads to bonds which do not rotate and thus
rigid structures and networks of low fractal dimension are formed. It appears
that for this geometry, a modest depletion attraction
can drive  the system into a regime similar to diffusion-limited 
cluster aggregation (DLCA) where the bonding is irreversible.
Inspired by the remarkable effect of particle geometry, we develop a
model which captures the essential geometrical properties of the colloidal rocks
in the form of dodecahedral clusters of 20 spheres, as shown in Fig.
\ref{figPix}(e). We further explore the effect of particle geometry at high density and find that 
the colloidal rocks are a quasi one-component glass-former.

\section{Experimental}

Faceted polytetrahedral rock-shaped particles as shown in Fig. \ref{figPix}(a) were produced by 
modifying a synthesis to produce poly-methyl-methacrylate (PMMA) colloids stablised with poly-hydroxy stearic acid \cite{dullens2003}. Details of the synthesis  
will be presented elsewhere.
The particle size is expressed as $\sigma_{r}=3.5$ $\mu$m, the longest dimension of each rock 
in images such as Fig. \ref{figPix}(a). It is hard to determine
the polydispersity for these particles, which have both shape and
some size polydispersity. However, inspection of SEM images suggests
that the size polydispersity is $\sim5$\%. The colloidal rocks were
labelled with rhodamine iso-thiocynate fluorescent dye. To investigate the change 
induced by anisotropy we also used
spherical PMMA particles, of diameter $\sigma_s=2.4$ $\mu$m with 4\% polydispersity 
determined by static light scattering.

The colloids were dispersed in a density- and refractive index matching
mixture of cis-decalin and cyclohexyl bromide (CHB), to which was added
4 mMol of tetrabutyl ammonium bromide salt to screen electrostatic
interactions \cite{royall2008g}. Polystyrene polymer was added to
induce depletion attractions. The polymer used was $M_{w}=2.1\times10^{7}$ g/mo
in the case of the rocks and $M_{w}=3.1\times10^{7}$  g/mol  for the spheres resulting in 
polymer radii of gyration of $R_{g}=180$ nm and $R_{g}=220$ nm
\cite{royall2007} and a polymer-colloid size ratio $2R_{g}/\sigma$ of $0.11$
and $0.18$, respectively. With the exception of the modest change in
interaction range, the experimental conditions were identical for both
rocks ($r$) and spheres ($s$). The colloid volume fraction
$\phi_{r}=\phi_{s}=0.05$ was determined by weighing out the samples. 
We track the colloid coordinates in 3D using confocal microscopy images. 
The rocks are sufficiently large with an intensity maximum at the centre of the
particle that algorithms developed for spheres \cite{royall2003} identify the rock centres 
effectively with an error $\lesssim 100$ nm.
%We confirmed that rock centres were successfully tracked
Particles were defined as bonded if their centres 
lay within the interaction range $\sigma+2R_g$, i.e. $3.9$ $\mu$m and $2.8$
$\mu$m for the rocks and spheres respectively. Moderate changes
in this bond length had no significant impact on our results.

\section{Results and Discussion}

Geometry introduces two key differences between the rocks and
spheres relevant to depletion attractions as indicated in Fig. \ref{figPix}(b). 
Firstly, for rocks the overlap of excluded volume, which drives the
depletion interaction, is much reduced compared to spheres. 
This is due to the concave shape of the rocks which means that 
much of the volume from which the polymers are excluded cannot overlap due to 
the approach of two rocks. This is shown by the pink regions in Fig.  \ref{figPix}(b) which denote
excluded volume, while the black is the colloidal particle. Blue denotes regions where
overlap of excluded volume is prevented due to the concave faces.
Thus we expect
that more polymer is required in the case of the rocks, for a given degree of
attraction. Secondly, the faces of the rocks lead to multiple overlap
zones as shown by the overlapping pink regions in Fig. \ref{figPix}(b). 
These impose an energetic penalty to rotation, which is absent
in the case of spheres. Both points were confirmed by calculations with
model rocks and ideal polymer.

\begin{figure*}
\includegraphics[width=78mm]{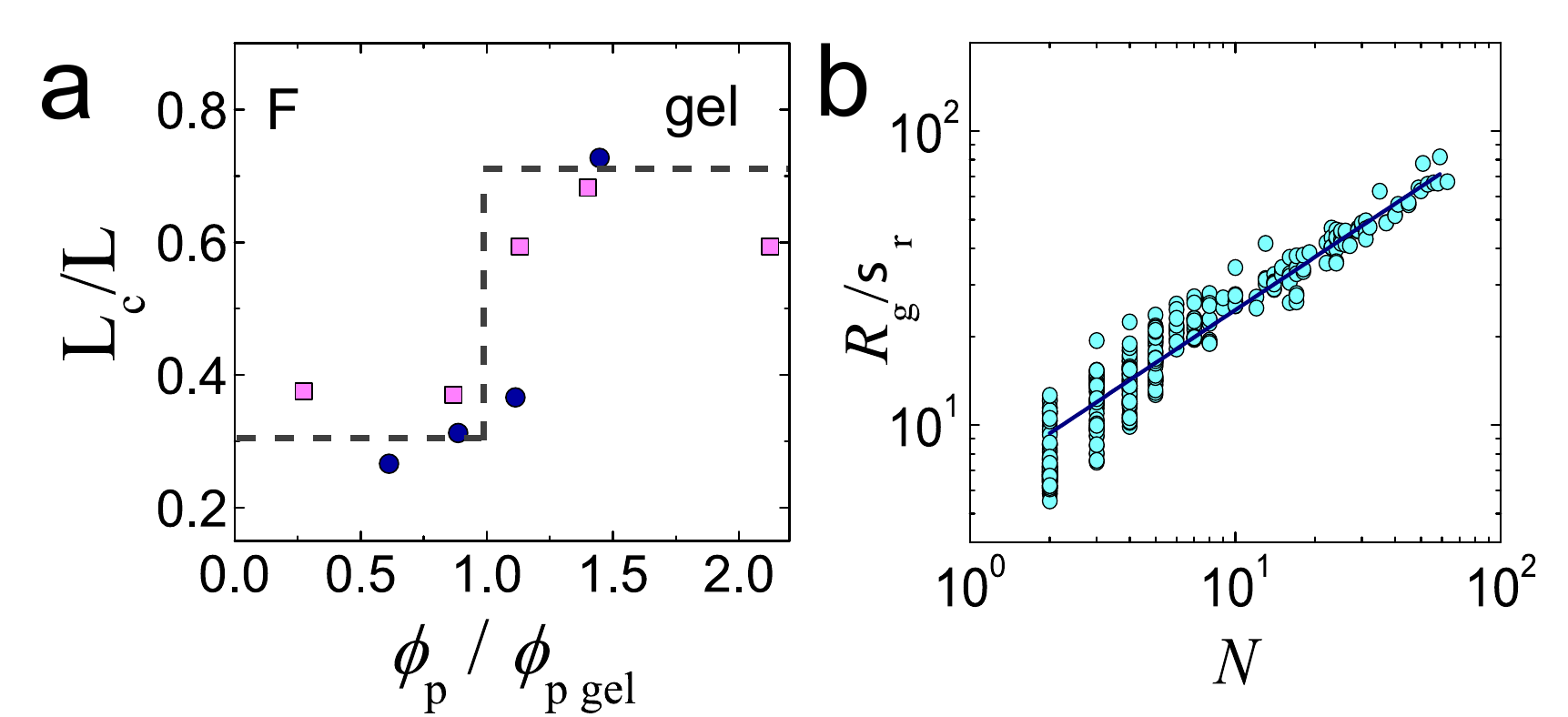}
\includegraphics[width=82mm]{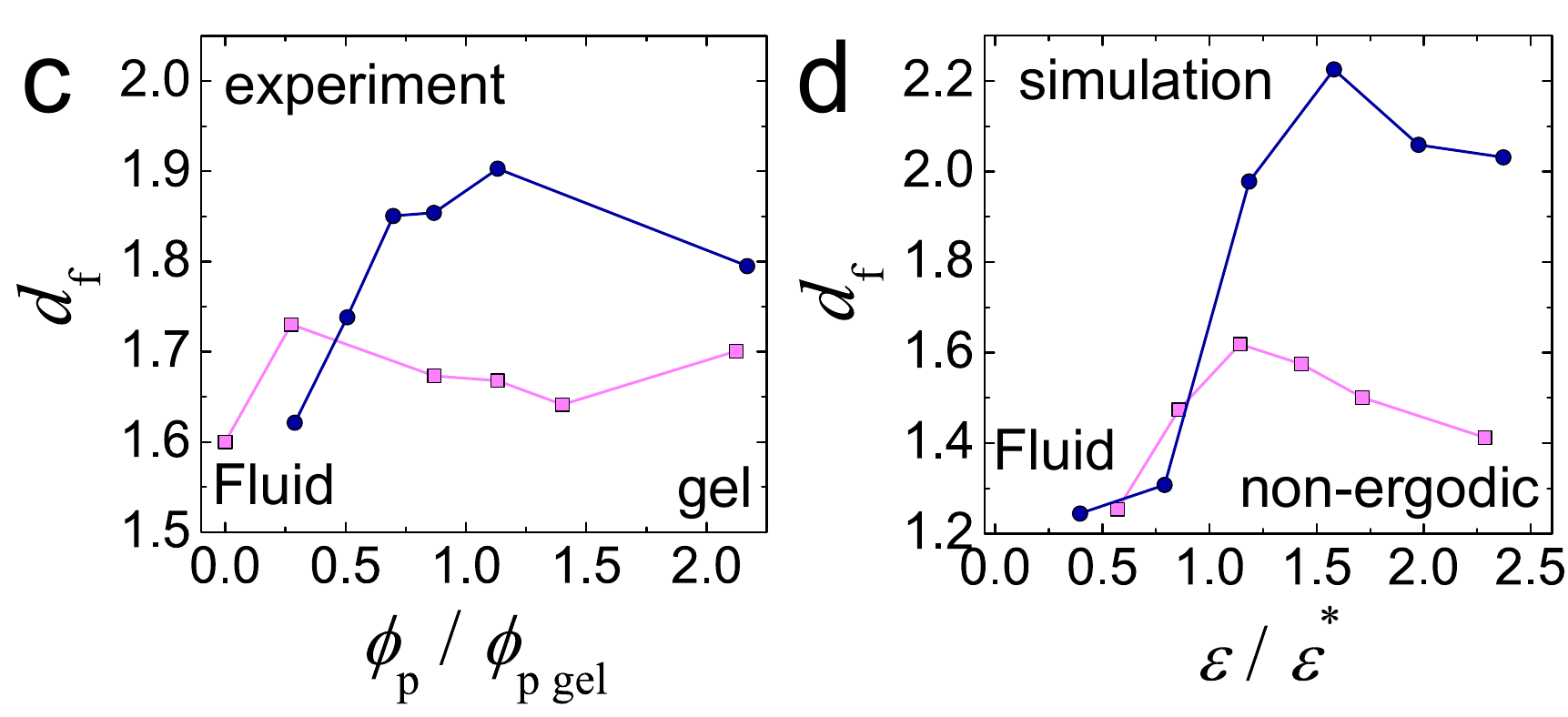}
\caption{(color online) Percolation and fractal dimension. (a) Size
of largest connected region $L_{c}$ scaled by image size $L$. Dashed line is a guide to the eye. (b) Radius of gyration of clusters
as a function of the number of particles in the cluster plotted to yield the fractal dimension
$d_{f}$ (experimental data from rocks). $d_{f}$ in experiments (c) and in
simulations (d) for isolated clusters with $\phi_{r}=\phi_{s}=0.0125$. 
In (a,c,d) the rocks are denoted by pink squares
$\square$ and the spheres by circles $\bullet$.}
\label{figPerky} 
\end{figure*}

Confocal microscopy images of gels of rocks and spheres are shown
in Fig. \ref{figPix}(c) and (d). The difference is striking:
rocks form one-particle wide chains while spheres form densely
packed structures. This provides direct evidence for a different bonding
scenario in the two species, as indicated in Fig. \ref{figPix}(b).
We identify gelation with a percolating network of long-lived bonds. 
Gel dynamics are shown in supplementary movie 1. In finite sized microscope
images, it can be hard to directly measure percolation. We therefore determine percolation through the ratio $L_{c}/L$ as shown
in Fig \ref{figPerky}(a), where $L_{c}$ is the length of the largest
connected cluster and $L$ is the image size. Gelation is thus identified
with a rapid rise in the ratio $L_{c}/L$. The polymer \emph{mass} fraction
for gelation $c_{p}^{gel}$ in the case of spheres is 
$4.1\pm0.5\times10^{-4}$ and $2.0\pm0.3\times10^{-3}$ for
the rocks, which corresponds to a \emph{volume} fraction
$\phi_{p}^{gel}$ of $0.42\pm0.05$ for the spheres and $2.3\pm0.2$ for the
rocks. Here the polymer volume fraction is defined $\phi_{p}=4 \pi \rho_p R_g^3 /4$
where $\rho_p$ is polymer number density.  We attribute this fivefold change in
polymer volume fraction required for gelation to the much reduced
overlap volume in the case of the rocks. Effects due to different polymer-colloid size
ratios and polymer non-ideality were estimated to be small \cite{louis2002}.

\begin{figure}
\includegraphics[width=40mm]{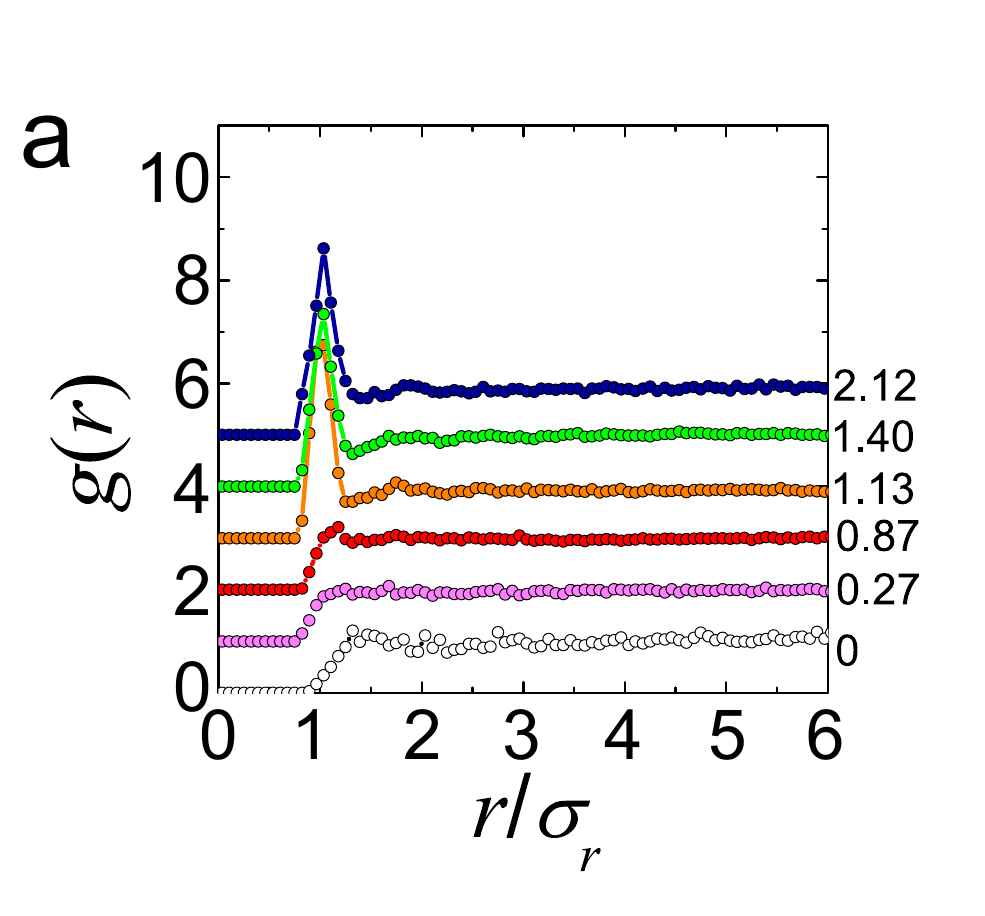}
\includegraphics[width=40mm]{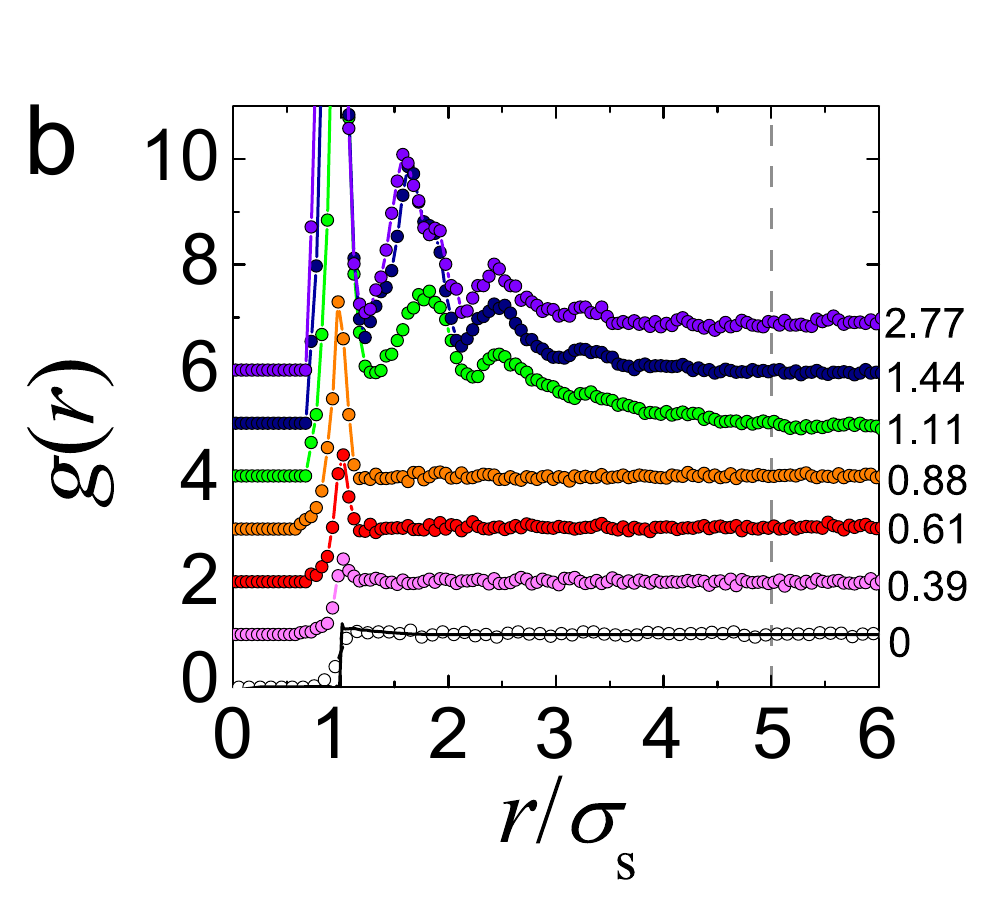}
\includegraphics[width=80mm]{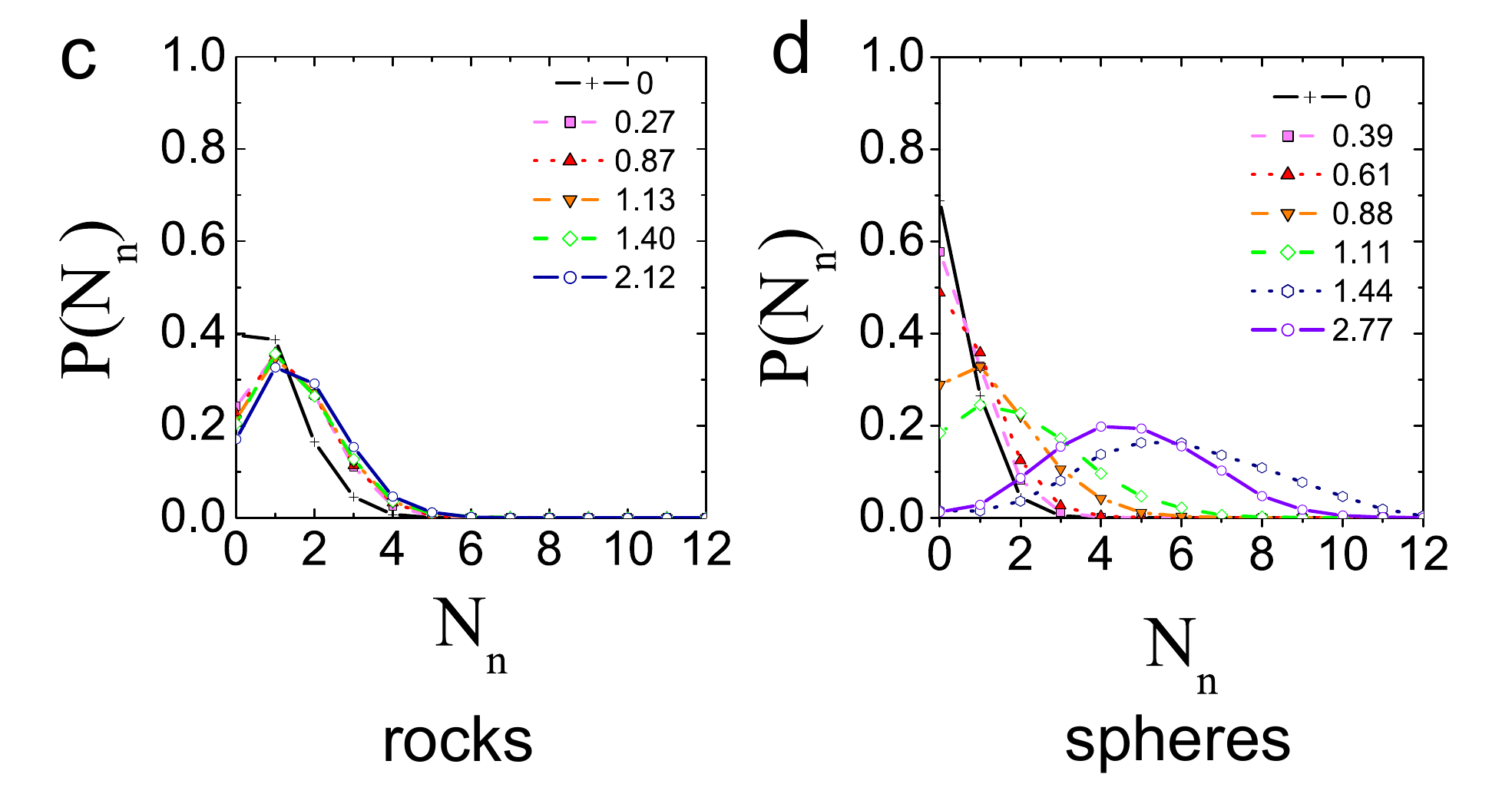}
\caption{(color online) Radial distribution functions $g(r)$ for rocks (a)
and spheres (b), respectively. Dashed line in (b) denotes the approximate
lengthscale of incipient phase separation for the spheres.
Data are shifted for clarity. Distributions of the number of neighbours $P(N_n)$ for rocks (c)
and spheres (d), respectively. Labels denote $c_{p}/c_{p}^{gel}$.}
\label{figGNN}
\end{figure}

The difference between the rocks and spheres in
the structure at the particle level is also striking, as shown by the radial distribution
functions in Fig. \ref{figGNN}(a) and (b). Data for spheres has been compared against
computer simulation for hard spheres in \cite{royall2007}, showing excellent agreement.
The rocks show similar behaviour, indicating that the colloids are stable against
aggregation. At higher polymer concentration, although both species
show a strong increase in the first peak upon gelation,
in the case of spheres, the onset of gelation ($c_p/c^{gel}_p \geq 1$)  
coincides with
 $g(r)>1$ for $r<5\sigma_s$,  indicating
clustering/dense regions in agreement with Fig. \ref{figPix}(d).
This behaviour is entirely absent in the case of the rocks [Fig.
\ref{figPix}(c)]. In other words, for spheres, phase separation has
already begun before gelation sets in: gelation is driven by
the arrest due to the development of the dense phase \cite{lu2008},
once it has formed on a lengthscale of several particles.
The rocks show no signs of locally dense regions: gelation is
driven soley by rigid bonds. Similar conclusions may be reached by considering
the number of neighbours of a given particle [Fig. \ref{figGNN}(c,d)]. 
In the case of spheres, the onset of phase separation (gelation) leads to a considerable
change in the number of neighbours, indicating locally dense regions. 
Rocks on the other hand show no such indication of phase separation.
In other words, rocks have a reduced valency induced by their geometry
with some similarity to patchy particles \cite{bianchi2006}.

This bonding and local structure has
significant implications for the network formation. To quantify the structure,
we use a local `fractal dimension', $d_f$, for finite assemblies of
particles, which is defined as $R_{g}^c\propto N^{1/d_{f}}$ where $R_g^c$ is the
radius of gyration of a connected region of $N$ particles, as shown in 
Fig. \ref{figPerky}(b). At the low packing fractions we consider, the fluid phase forms
clusters, whose $d_f$ is found in the same way.
We plot the resulting $d_{f}$ in Fig. \ref{figPerky}(c). Spheres
and rocks show an increase in $d_{f}$ upon increasing the polymer
concentration, up to gelation, after which non-ergodicity leads to
a failure to relax locally and a decrease in $d_{f}$, as
found previously \cite{ohtsuka2008}. However,
the fractal dimension of rock clusters and gels is significantly lower than
that of spheres. This is consistent with the idea that the sphere-based
clusters have already begun to condense (leading to an increase
in $d_{f}$), while the rock clusters show no signs of such condensation.
We note that for the rock gels, the measured fractal dimension is
even less than that expected for DLCA
 where $d_{f}=1.8$. This is surprising: DLCA assumes unbreakable
bonds, which is entirely different from the weak depletion-induced gelation
expected in colloid-polymer mixtures. 

We have argued that the only relevant difference between the rocks and spheres
lies in their shapes and now enquire as to the geometrical origins
of the rock behaviour. Unlike other approaches to produce networks with low
fractal dimension which rely on a few sticky patches on each
particle to reduce the number of bonds \cite{bianchi2006},
here we consider only the particle geometry with no restriction
on the number of neighbours a particle can have, except the steric limitation which also applies to particle with a spherically symmetric interaction. We introduce a model based on dodecahedral assemblies of
20 overlapping spheres whose centres lie on a sphere, as shown in
Fig. \ref{figPix}(e). This model was choosen as a simple representation
of the rock shape, as it is reasonably resistant to ordered packing, and the faces should mimic
the rigid bonding of the rocks. These (small) spheres have a diameter of 
$2/7 \sigma_c$, where $\sigma_{c}$ is the diameter of the circumscribed 
sphere. The volume of a set of overlaping spheres can be calculated exactly
by e.g. the Connolly algorithm \cite{connolly1985} and is 
approximately 0.416 $\sigma_c^3$ for these model rocks.

For short ranged attractions, the detailed form
of the interaction potential is irrelevant for many properties \cite{noro2000},
so we treat the attractions that result from the addition of polymer
with a square well of width $0.042\sigma_{c}$ and depth
$\varepsilon$. We also carry out
simulations of spheres of diameter $\sigma_{s}$ interacting via the
square well potential of width $0.042\sigma_{s}$. Using the extended law of corresponding states, \cite{noro2000}, this maps to
a polymer-colloid size ratio of $1/7$, comparable to the experiments.
Here we focus on the geometric properties, rather than the dynamics: 
our purpose is to demonstrate the qualitatively
different nature of aggregation of spheres and rocks. We therefore
use Monte Carlo (MC) simulations in the canonical ensemble with 512 particles.
Simulations with more (2048) particles
showed minimal differences, as for the state points we consider.
We use a short
($0.04\sigma$) and long ($\sigma$) translational step for both spheres and rocks.
The former enables local restructuring of bonded particles, the latter
accelerates aggregation. For the rocks, 
small ($0.01$ rad) and large ($0.2$ rad) rotational moves
are also used. Each simulation is `equilibrated' for $10^{4}$ MC sweeps prior to sampling. 
Each sweep involves attempted moves of all types. Simulations are repeated at least six times.

In Fig. \ref{figPerky}(d) we plot fractal dimensions for simulations of 
rocks and spheres for packing fraction $\phi_{r}=\phi_{s}=0.0125$
(the experiments are carried out at $\phi_{r}=\phi_{s}=0.05$) \cite{simComment}. 
The spheres exhibit
similar behaviour as in the experiments, with $d_{f}\sim2$ when the
system falls out of equilibrium on the simulation timescale. This we take as the strength of attraction associated with spinodal decomposition \cite{lu2008}, which
corresponds to $\beta \varepsilon^* \approx 2.53$, noting that for such short-ranged
interactions, the spinodal line depends rather weakly on $\phi$.
For the rocks, we identify $\varepsilon^* \approx 1.75$ $k_BT$
with the onset of a regime where the potential energy decreases continuously.
Both rock and sphere $d_f$ are lower than the experiments at low attraction, which may be related to the change in $\phi$.

The simulations clearly show that the rock clusters have a fractal dimension which is less than 
that of the spheres [Figs. \ref{figPerky}(c,d)]. In fact $d_f \approx 1.5$, even less 
than in the experiments.  Although the simulation results are presented 
for a lower density the agreement between experiment and 
simulation adds further weight to our hypothesis that (\emph{i}) the rocks have an intrinsically
different bonding mechanism compared to spheres driven by geometry
and (\emph{ii}) that their behaviour is qualitatively captured by the dodecahedral clusters. 
Since the rocks can bond to as many neighbours as
they have faces, they they can form dense phases and therefore, at equilibrium dense and dilute phases coexist so the rock gels should be metastable.

The fractal dimension of rock clusters, both in experiments with percolating networks 
and simulations of isolated clusters is low, $\sim 1.7$ and $\sim 1.5$ 
respectively, lower even than DLCA.
Cates \emph{et.al.} \cite{cates2004}, suggested
that similar behaviour might be found in spheres, if they were suddenly -- and deeply --  quenched to the regime where DLCA dominates. The rock
geometry may be interpreted as exhibiting similar behaviour
with much weaker attractions.

\begin{figure}
\includegraphics[width=40mm]{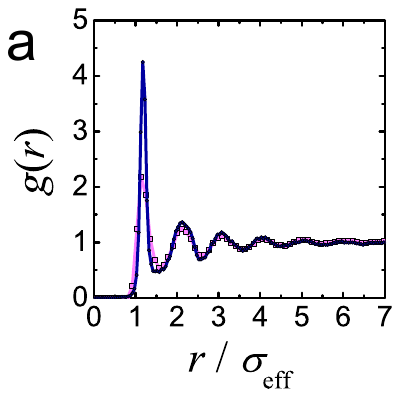}
\includegraphics[width=40mm]{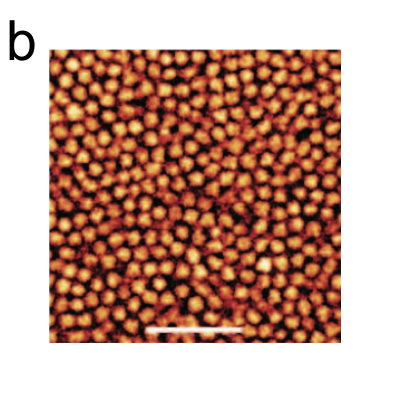}
\caption{(color online) (a) Mapping $g(r)$ of rocks at $\phi_{r}=0.48$ (pink
squares) and $g(r)$ of spheres at $\phi_{s}=0.54$ (black circles) from
experiments. $\sigma_{\text{eff}}$ denotes $\sigma_{r}$  and $\sigma_{s}$  for rocks
and spheres respectively. We estimate the mapping
by comparing the higher-order peaks. At the time of measurement the
metastable sphere fluid showed no signs of crystallisation.
(b) Confocal image of rock glassy state. Bar=10 $\mu$m.\label{figDenseG}}
\end{figure}

Finally, we consider the high density behaviour. In experiments
without polymer, rocks did not crystallise at a packing fraction
$\phi_{r}=0.46$ on a timescale of one week (Fig. \ref{figDenseG}). 
This is reasonable, as it has recently been shown that
only a limited degree of asphericity is required to suppress crystallisation
\cite{miller2010}.
Supplementary movie 2 shows that rocks at $\phi_{r}=0.46$
exhibit glassy dynamics, with displacement hindered by their neighbours.
We find that the radial distribution function of dense amorphous
packings of rocks at $\phi_{r}=0.46$ has peak locations and a decay of
the oscillations similar to the $g(r)$ of spheres at $\phi_{s}=0.54$
[Fig. \ref{figDenseG}(a)].  
This indicates
that the longer-range structure is similar, suggesting an effective packing fraction
for the rocks higher than $\phi_{r}=0.46$, likely due to the concave geometry.
The failure of the rocks to crystallise indicates that they are a `quasi one-component glass
former'.

\section{Conclusions}

We have developed an experimental model system of anisotropic colloidal
rocks whose interactions can be tuned and which can be
visualized directly in 3D. Due to their shape, the bonding introduced via the depletion
attraction is profoundly different to spheres: the bonds formed between
rocks are rigid against particle rotation. The frustration induced by this change in geometry profoundly influences the kinetic pathway:
rocks form open networks of low fractal dimension. We presume that the rock
gels are, like gels of spheres, ultimately metastable to phase separation,
but expect that the rock gels should be very long-lived. We demonstrated that geometry is the dominant driver for this behaviour by introducing a model of dodecahedral clusters which captures the essence of the colloidal
rocks. At higher density 
the experimental system forms a quasi-one-component glass. 

Our work opens up a single particle level approach to tackle
important problems such as the role of friction and surface roughness in colloidal and nanoparticle
 processing and jamming. We hope to stimulate further experiments,
for example the use of external fields such as shear, along with theoretical and simulation
work which may fully determine the phase diagram, and the role of particle geometry in kinetic trapping and its interplay with hydrodynamic interactions. Self assembly of nanodevices requires avoidance of kinetic trapping \cite{whitesides2002}; here we have shown the potential of direct visualisation as a means to develop control of self-assembly which may yield insight into nano-assembly.

\noindent \textbf{Acknowledgments} R. Rice and CPR gratefully acknowledge the Royal
Society for financial support. The authors would like to thank Doug Ashton, Paul
Bartlett, Patrick Charbonneau, Bob Evans, Rob Jack and John Russo for helpful discussions. 

%\bibliography{rox}

\begin{thebibliography}{24}
\bibitem[{sim()}]{simComment}
\bibinfo{note}{The simulations are carried out a lower packing fraction to
  avoid the effects of percolation which limit the statistics gathered in our
  calculation of df.}
\end{thebibliography}
%\bibliographystyle{apsrev}

\end{document}